# Pressure-induced superconductivity in SnSb$_2$Te$_4$

*Peng Song[a,b], Ryo Matsumoto[a,b], Zhufeng Hou[c], Shintaro Adachi[a], Hiroshi Hara[a,b], Yoshito Saito[a,b], P. B. Castro[a,b], Hiroyuki Takeya[a], and Yoshihiko Takano[a,b]*

[a]*International Center for Materials Nanoarchitectonics (MANA),*

*National Institute for Materials Science, 1-2-1 Sengen, Tsukuba, Ibaraki 305-0047, Japan*

[b]*University of Tsukuba, 1-1-1 Tennodai, Tsukuba, Ibaraki 305-8577, Japan*

[c]*Fujian Institute of Research on the Structure of Matter (FJIRSM), Chinese Academy of Sciences,*

*Fuzhou, 350002 Fujian, People's Republic of China*

**Abstract**

We report the discovery of a new superconductor from phase change materials $SnSb_2Te_4$. Single crystals of $SnSb_2Te_4$ were grown using a conventional melting-growth method. The sample resistance under pressure was measured using an originally designed diamond anvil cell with boron-doped diamond electrodes. The pressure dependence of the resistance has been measured up to 32.6 GPa. The superconducting transition of $SnSb_2Te_4$ appeared at 2.1 K($T_c^{onset}$) under 8.1 GPa, which was further increased with applied pressure to a maximum onset transition temperature 7.4K under 32.6 GPa.

Phase change materials (PCMs) are distinguished by their excellent optical and electrical properties and the fast switching between amorphous and crystalline phase and thus are widely used in optics and data storage.[1][2] Most PCMs are composed of group IV, V and VI elements, such as $GeSb_2Te_4$ and $GeSb_2Te_5$, and the Fermi level of these compounds at ground states is often located inside the band gap, exhibiting semiconducting properties.[3] Recently, superconducting properties of these binary or ternary PCMs, such as SnTe[4], $Sb_2Te_3$[5], $GeSb_2Te_4$[6] have been found whenever pressure is applied or carriers are introduced. First-principles calculations have shown that the high-pressure phase (Pm-3m) of SnTe is metallic and has a very flat band near the Fermi level[7], which is favorable to the formation of the cooper pairs.[8] The superconductivity in $Sb_2Te_3$ indicates that the crystal structure of the compound remains stable at lower pressure and its superconductivity is topologically related.[9] Here we focus on the PCM $SnSb_2Te_4$, which can be considered a mixture of SnTe and $Sb_2Te_3$, so it is of great interest to check whether this ternary compound would exhibit superconductivity under applied pressure or not. Herein we focus on the pressure-dependent structure and electronic properties under pressure of $SnSb_2Te_4$.

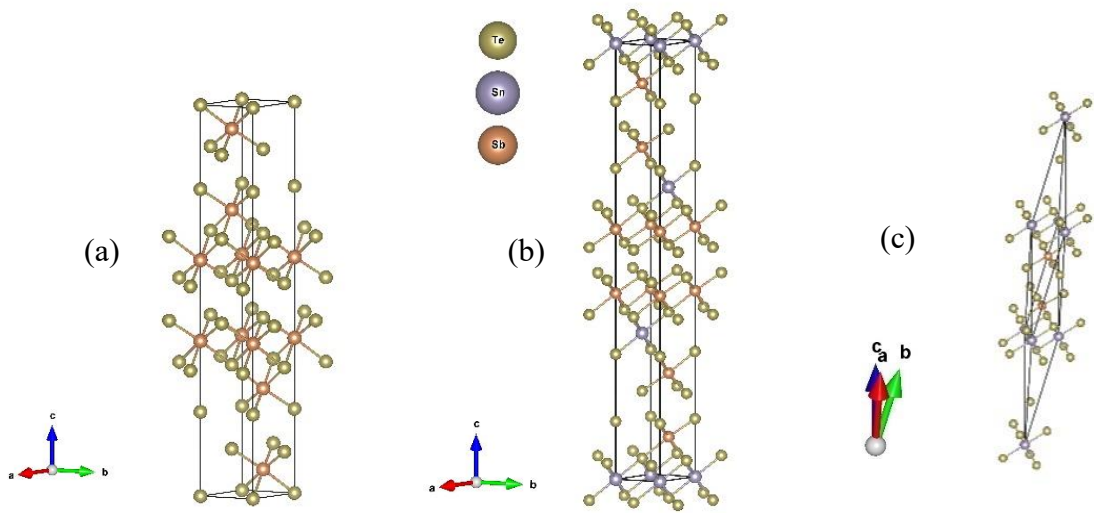

Figure 1(a) parent compound of $Sb_2Te_3$, (b) crystal structure of $SnSb_2Te_4$. (c) Primitive cell of $SnSb_2Te_4$

Figure 1 shows crystal structure of (a) parent compound $Sb_2Te_3$, (b) of $SnSb_2Te_4$, (c) of its primitive cell. The figures were drawn using VESTA[10]. Brown, green yellow, and gray balls represents Sb, Te, and Sn, respectively. $SnSb_2Te_4$ has a trigonal structure with space group R-3m(lattice constants of a = b = 4.3158 Å and c = 41.6574 Å) at ambient pressure.[11][12] The intermediate layers of $Sb_2Te_3$ and $SnSb_2Te_4$ are almost identical, and each Sb atom is coordinated with four Te atoms. The structure difference between $SnSb_2Te_4$ and $Sb_2Te_3$ is that the central Te2 layer in $Sb_2Te_3$ is replaced by the Te2-Sn-Te2 layer in $SnSb_2Te_4$. This ternary compound is weakly bonded by van der Waals interaction between Te layers.[13][14] By a certain heat treatment, the $SnSb_2Te_4$ also transform into a NaCl-type FCC structure[15].

The atomic structure and electronic structure of $SnSb_2Te_4$ were calculated using the projector augmented wave (PAW) method, as implemented in the Quantum ESPRESSO software package.[16][17] The generalized gradient approximation(GGA) of Perdew-Burke-Erzerhof (PBE)[18] was used to describe the exchange-correlation functional. The inner 4d-electrons in Sn and Sb are treated as valence electrons. A 10 x 10 x 10 k-grid was employed for the k-point sampling in the first Brillouin zone and the kinetic energy cutoffs for the expansion of electronic wave function was set to 74 Ry. The atomic positions and lattice parameters were relaxed using the Broyden-Fletcher-Goldfard-Shanno(BFGS) algorithm. In the density of states (DOS), a k-mesh size of 20 x 20 x 20 was used. The Crystal Orbital Hamiltonian Population (COHP)[19] and its energy integral (ICOHP) were used for bonding state analysis, implemented in LOBSTER.[20]

Figure 2 shows the calculated band structure, density of states (DOS), and COHP of $SnSb_2Te_4$ at ambient pressure. We also present the band structure of $SnSb_2Te_4$ (Fig.2(d)) under high pressure of 10 GPa for a comparison. We can see that this compound is a semiconductor with a narrow band-gap

(~0.26 eV obtained by the present GGA-PBE calculations) at the Z point. The energy bands near the top of valence bands and the bottom of conduction bands are both flat, and thus the valence and conduction band edges have high DOS. The orbital-decomposed DOS reveals that the conduction band of SnSb$_2$Te$_4$ is mainly composed of Sn-5p, Sb-5p and Te-5p states, and there is also an additional contribution from Te-5s states. The valence bands from -5 eV to -1 eV are mainly contributed by the hybridization between the Sn-(and Sb-) 5p and Te-5p orbitals, while the bands from -1 eV to the valence band maximum (VBM) are contributed by the hybridization between the Sn-(and Sb-) 5s states between the Te 5p orbitals. Figure 2(c) shows the calculated -pCOHP for the nearest-neighboring atom pairs of Sn-Te and Sb-Te, in which the positive (negative) of -pCOHP values correspond to the bonding (anti-bonding) characteristics. Therefore, we can see that the Sn-5p (and Sb-5p) and Te-5p orbitals form the bonding states in the energy range from -5 eV to -1 eV, while the Sn-5s (Sb-5s) and Te-5p orbitals form the anti-bonding states in the energy range from -1 eV to the VBM. The anti-bonding states, which come from the hybridization between the Sn-5p (Sb-5p) and Te-5p, appear in the vicinity of conduction band bottom (CBM).

Compared with the Sb-Te bond length in Sb$_2$Te$_3$, the Sn-Te and Sb-Te bond lengths in SnSb$_2$Te$_4$ have negligible change (0.02 Å at most). The characteristics of chemical bonding between Sb-Te in two compounds are very similar. It is noted that the energy levels of Sn 5s (5p) are shallower than those of Sb. Therefore, SnSb$_2$Te$_4$ has a smaller energy gap than Sb$_2$Te$_3$. Figure 2(d) shows the band structure of SnSb$_2$Te$_4$ under pressure of 10 GPa. It can be seen that the Fermi level crosses the top valence bands, indicating the pressure-induced a decrease of the band gap of SnSb$_2$Te$_4$ and a transition to metallic behavior. As mentioned above, the anti-bonding characteristic appears both in the vicinity of VBM and CBM of SnSb$_2$Te4. If the crystal structure does not transform into a different phase under high pressure, the applied pressure will usually lead to a decrease of bond lengths between atoms. For SnSb$_2$Te$_4$, the applied pressure results in the shrinking of Sn-Te and Sb-Te bond lengths and thus the anti-bonding states near the VBM and CBM would both shift toward high energy. This might be the reason why the band gap of SnSb$_2$Te$_4$ decreases as the applied high pressure. The similar trend was also found in our previous studies for SnBi$_2$Se$_4$[21], PbBi$_2$Te$_4$[22], and AgIn$_5$Se$_8$[23].

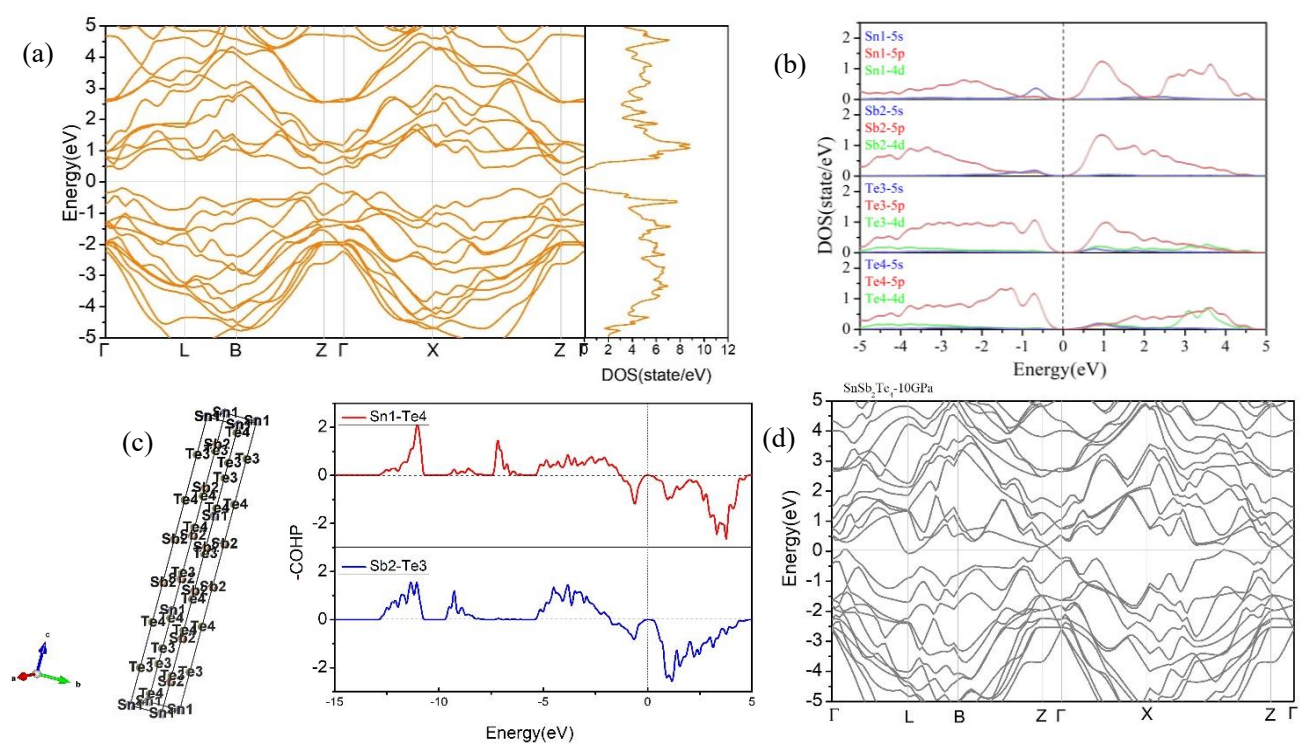

Fig 2. (a) band structure of SnSb$_2$Te$_4$ and electron density of state (DOS) of SnSb$_2$Te$_4$ (b) projected density of state(PDOS) (c) the calculated crystal orbital Hamilton population for the SnSb$_2$Te$_4$ at zero pressure (d) the comparative band structure at 10 GPa

Single crystals of SnSb$_2$Te$_4$ were grown using a conventional melting method. Sn (99.9%, powder), Sb (99.99%, powder), Te (99.9%, grain), combined in stoichiometric ratios in an evacuated silica tube. Afterward, the tube was put into a furnace and heated up to 1010K for 10h. The samples were then slowly cooled to 873K at a rate of 9.1 K h$^{-1}$ and held for 24h. Powder X-ray diffraction was conducted by Mini Flex 600 (Rigaku). Pressure effect was measured by Boron-doped diamond anvil cells (DAC)[24]. Figure 3 shows the powder XRD pattern of SnSb$_2$Te$_4$. The peaks were corresponding to a

R-3m(H) structure of SnSb$_2$Te$_4$, with a lattice constant of a=b=4.304(1) Å and c = 41.739(3) Å. Sn Sb and Te are very uniformly distributed in this material. Using EDS to analyze its composition, the ratio of the material was shown as Sn$_{1.01}$Sb$_{1.98}$Te$_4$, which shows good agreements with nominal composition.

The temperature dependence of the resistivity for SnSb$_2$Te$_4$ are shown in Fig. 4, measured from

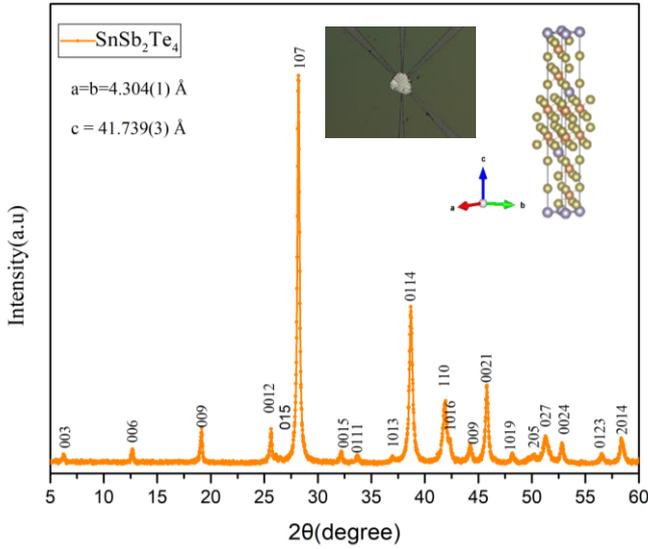
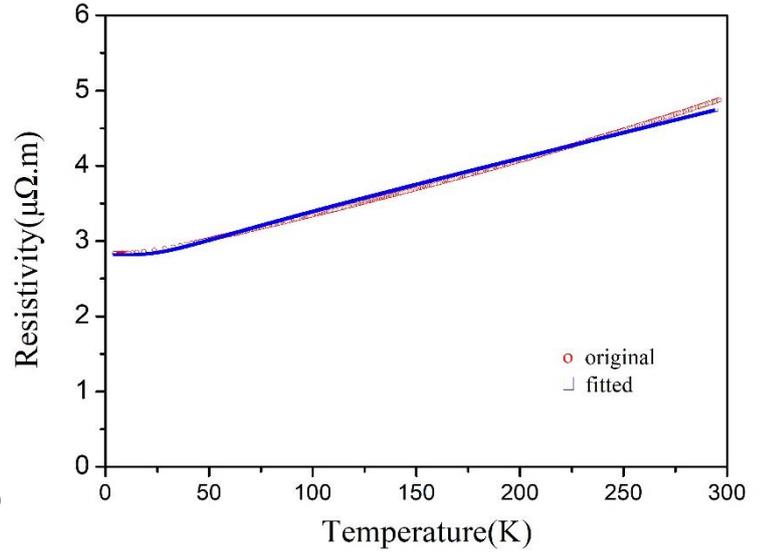

Fig. 3. Room temperature X-ray diffraction patterns of SnSb$_2$Te$_4$

Fig. 4. Temperature dependent resistivity of SnSb$_2$Te$_4$

4.5K to 296K. Fig. 4 shows the resistivity of SnSb$_2$Te$_4$ decreases with decreasing temperature which is in good agreement with metallic behavior. In non-magnetic metals, the temperature dependence of resistivity is mainly derived from electron-phonon interactions [25] and can be expressed by Bloch-Gruneissen formula.[26]

$$\rho(T) = \rho(0) + \alpha_{el-ph}(\frac{T}{\Theta_R})^5 \int_0^{\frac{\Theta_R}{T}} \frac{x^5}{(e^x - 1)(1 - e^{-x})} dx$$

Where ρ(0) is the residual resistivity due to defect scattering and is independent of temperature. $\Theta_R$ is the Debye temperature. α$_{el-ph}$ is a constant associated with the electron-phonon coupling constant λ. The blue line of Fig4(a) is the fitted equation from 6K to 200K and the Debye temperature can be obtained close to 320K, which is much larger the Sb$_2$Te$_3$($\Theta_R$ = 200K)[27] and SnTe ($\Theta_R$ = 165K)[28]. It is known from the BCS theory that the Debye temperature is positively correlated to the critical temperature $T_c$.[27] Therefore, we have reason to believe that SnSb$_2$Te$_4$ can exhibit superconducting properties under pressurized conditions.

Figure 5, shows the transport measurements at low temperature at pressures between 6.3GPa to 32.6GPa. Under these pressures, the relationship between resistance and temperature reveals significant metallic behavior. It is also apparent that superconductivity begins to occur at 8.1 GPa with $T_c^{onset} \sim$ 2K and the zero resistivity was observed at 10.2 GPa with $T_c^{zero} \sim$2.2K. At a further increase of pressure, in the range between 10.2 and 23 GPa, the superconducting transition temperature increases with increasing pressure, and $T_c^{zero}$ maintains a linear relationship with pressure. Figure 5(c) shows the superconducting phase diagram of the SnSb$_2$Te$_4$ single crystal. The resistance begins to drop sharply at 10.2 GPa. Moreover, starting from 10.2 GPa, there is also a linear relationship between resistance and pressure. At 32.6 GPa, the highest critical temperature $T_c^{onset}$ reaches 7.4K.

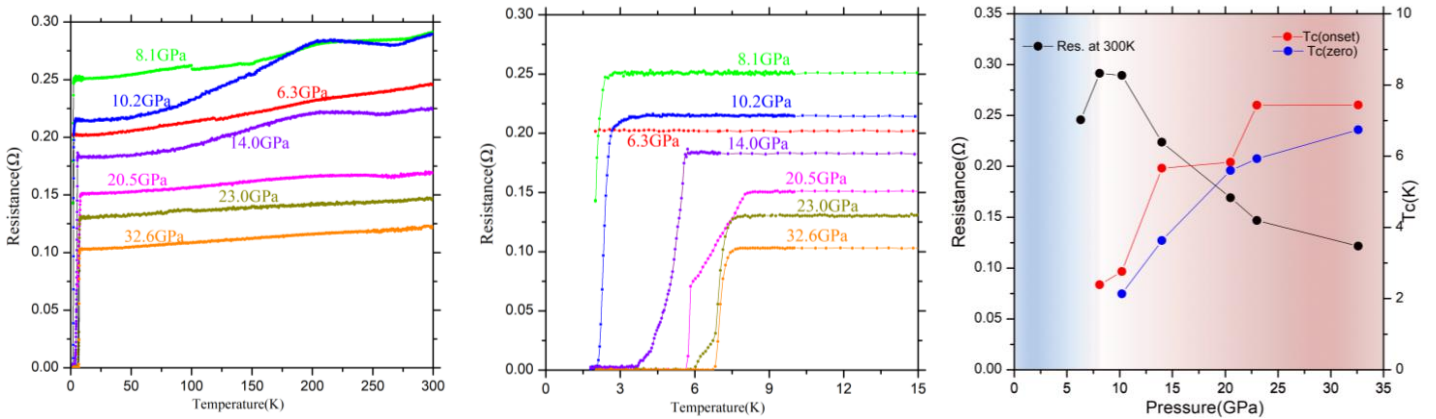

Figure 5(a) temperature dependence of resistance from 2K to 300K (b) temperature dependence of resistance from 2K to 15K (c) superconducting phase diagram

According the context of weaking-coupling Bardeen-Cooper-Schrieffer (BCS) superconductivity theory, $\kappa T_c = \hbar\omega_{ph}e^{-1/\lambda N(E_f)}$ , where $\kappa$ is the Boltzmann constant, $\hbar\omega_{ph}$ is a phonon energy, $\lambda$ is the electron-phonon coupling constant, and $N(E_f)$ is the density of state at the Fermi level. Therefore, increasing the density of states of the Fermi level contributes to the transition from a normal conductor to a superconductor. We have found that in the case of anti-bonded substances, the energy gap tends to decrease with increasing pressure when the pressure is not large. Since SnSb2Te4 is generally flat near the Fermi level, and an anti-bond state occurs in the bonding region. Therefore, under pressurized conditions, electron-phonon coupling is likely to occur, resulting in superconductivity. Since we did not obtain the XRD data under high pressure, the specific superconducting mechanism of SnSb2Te4 is still unknown.

**Conclusion**

We study the structure information of SnSb$_2$Te$_4$ via first-principles calculation. We found that SnSb$_2$Te$_4$ has anti-bond state at the valence and conduction band edges. We have noticed that many pressure-induced superconductor in our previous studies have this similar charactersitcs. The SnSb$_2$Te$_4$ undergoes a superconducting transition with $T_c^{onset} \sim 2K$ when a pressure of 8.1 GPa is applied. With further increase of pressure, the systems reaches a maximum of $T_c^{onset} \sim$ 7.4 K, around 32.6 GPa. And the relationship between pressure and Tc did not appear saturated.

**Acknowledgements**

This work was partly supported by JST CREST Grant No. JPMJCR16Q6, JST-Mirai Program Grant Number JPMJMI17A2, JSPS KAKENHI Grant Number JP17J05926, 19H02177, and the "Materials research by Information Integration" Initiative (MI2I) project of the Support Program for Starting Up Innovation Hub from JST. The computation in this study was performed on Numerical Materials Simulator at NIMS.